\documentclass[showpacs,preprintnumbers,10pt,twocolumn]{revtex4}%
\usepackage{amssymb}
\usepackage{amsfonts}
\usepackage{amsmath}
\usepackage{graphicx}
\usepackage{dcolumn}
\usepackage{bm}
\usepackage{revsymb}%
\begin{document}
\title{Comment on \textquotedblleft Topological Pumping in a Floquet-Bloch
Band\textquotedblright}
\author{Ricardo Chac\'{o}n$^{2}$ and Pedro J. Mart\'{\i}nez$^{1}$}
\affiliation{$^{1}$Departamento de F\'{\i}sica Aplicada, E.I.I., Universidad de
Extremadura, Apartado Postal 382, E-06006 Badajoz, Spain, and Instituto de
Computaci\'{o}n Cient\'{\i}fica Avanzada (ICCAEx), Universidad de Extremadura,
E-06006 Badajoz, Spain}
\affiliation{$^{2}$Departamento de F\'{\i}sica Aplicada, E.I.N.A., Universidad de Zaragoza,
E-50018 Zaragoza, Spain}
\date{\today}

\begin{abstract}

\end{abstract}
\maketitle

In a recent Letter [1], topological pumping in a Floquet-Bloch band using a
plain sinusoidal lattice potential and two-tone driving with relative phase
$\varphi$, amplitudes $K_{1}$ and $K_{2}$, and frequencies $\omega$ and
$2\omega$, respectively, is experimentally and numerically demonstrated. For a
fixed amplitude $K_{1}$, the authors obtained numerical estimates of the
critical parameters $K_{2c},\varphi_{c}$ for closing the\ \textit{s-p}
bandgap, thus optimizing directional motion of the center-of-mass position of
the atomic cloud of ultracold $^{40}$K fermionic atoms, while they found that
the efficiency of this topological transport seems to be quantized for
sufficiently long driving periods (adiabatic directed transport (DT)). The
authors do not provide any theoretical explanation either for these apparently
magical values of $K_{2c},\varphi_{c}$ or for the \textquotedblleft quantized
efficiency\textquotedblright\ in the adiabatic regime. In this Comment, we
argue how the theory of ratchet universality (RU) [2-4] provides a satisfactory
explanation of such experimental findings. Although the numerical results
discussed in [1] are based on a Floquet two-band model of \textit{s, p} bands,
the authors confirmed that the pump efficiency is the same when evaluated
using the full time-dependent Hamiltonian%
\begin{align}
H\left(  \tau\right)   &  =\frac{p^{2}}{2}+V\left(  x\right)  -F\left(
\tau\right)  x,\tag{1}\\
F\left(  \tau\right)   &  \equiv\frac{\hbar}{a\omega}\left[  K_{1}\cos\left(
\omega\tau\right)  +2K_{2}\cos\left(  2\omega\tau+\varphi\right)  \right]
,\tag{2}%
\end{align}
which characterizes the dynamics of the center-of-mass position of the atomic
cloud in the frame co-moving with the driven optical lattice (throughout this
Comment, we follow the notations and definitions used in the Letter). Clearly,
DT appears due to the de-symmetrization of Floquet eigenstates when the
generalized parity symmetry and the generalized time-reversal symmetry are
violated, i.e., when $K_{1},K_{2}\neq0$ and $\varphi\neq0,\pm\pi$,
respectively [5], while the theory of RU predicts that there exists a
universal force waveform that optimally enhances DT by critically breaking
such symmetries. This theory has explained previous experimental results
concerning DT of fluxons in uniform annular Josephson junctions in one case
and of cold atoms in dissipative optical lattices in another, both driven by
biharmonic fields [6], and refers to the criticality scenario that emerges
when the aforementioned two symmetries are broken, regardless of the
nature of the evolution equation in which the breaking of such symmetries
results in DT. Specifically, the four equivalent expressions of the biharmonic
universal excitation [1] are given by $\cos\left(  \omega\tau\right)
\pm(1/2)\sin\left(  2\omega\tau\right)  $, $\sin\left(  \omega\tau\right)
\pm(1/2)\sin\left(  2\omega\tau\right)  $, i.e., the optimal values predicted
from RU for the force $F\left(  \tau\right)  $ [Eq.~(2)] to yield optimal
enhancement of DT of ultracold $^{40}$K fermionic atoms are
\begin{equation}
K_{1opt}=4K_{2opt},\varphi_{opt}=\pm\pi/2,\tag{3}%
\end{equation}
where the two signs $\pm$ correspond to DT in opposite directions. For the
value $K_{1}=0.8$ considered in [1], Eq.~(3) predicts $K_{2opt}=0.2$ that is
very close to the critical value $K_{2c}=0.22$ stated in [1]. In fact, the
authors considered the exact value $K_{2c}=K_{2opt}=0.2$ for the
parametrization of all pumping orbits (cf. Fig.~3 and Supplemental Material in
Ref.~[1]) and found $\varphi_{c}=\varphi_{opt}=\pm\pi/2$, as predicted from
RU. Additionally, the efficiency of the DT is predicted from RU to reach a
constant value that is independent of the pump period when this is the largest
timescale of the problem, i.e., when $\pi/\omega$ is sufficiently larger than
the inverse of the minimal gap size (adiabatic transport regime), providing
thereby an explanation for the \textquotedblleft quantized
efficiency\textquotedblright\ indicated by the authors (cf. Figs.~4 and S1 in
the Supplemental Material [1]). Otherwise, there is expected from RU a gradual
decrease of the efficiency as $2\omega$ is increased from the adiabatic limit,
i.e., as the relevant symmetries are gradually restored. This phenomenon of
competing timescales leads to the $2\omega$-force losing ratchet
effectiveness, but without deactivating the degree-of-symmetry-breaking
mechanism [2,3], thus explaining the gradual decrease of the efficiency when
$2\pi/\omega\rightarrow0$ (cf. Fig.~4 in Ref.~[1]). It is worth mentioning
that the same gradual decrease has been observed previously (cf. Fig.~8 in Ref.~[5]
and Fig.~5 in Ref.~[7]). To conclude, the results of Ref.~[1] show a fruitful
cooperation between the Berry phase for energy bands and the ratchet universal
excitation at optimally inducing charge pumping in driven lattice systems.

R.C. acknowledges financial support from the Junta de Extremadura (JEx, Spain)
through Project No. GR21012 cofinanced by FEDER funds. P.J.M. acknowledges
Grant No. PID2020-113582GB-100 funded by Spanish
MCIN/AEI/10.13039/501100011033 and by \textquotedblleft ERFD A way of making
Europe\textquotedblright.

\end{document}